\documentclass[letterpaper,titlepage,11pt]{article}
\usepackage{hyperref}

\usepackage{amssymb,amsmath,amsfonts}

\def\clock{{\count0=\time
           \divide\count0 60
           \ifnum\count0<10 0\fi\the\count0
           \multiply\count0 -60 \advance\count0 \time
           :\ifnum\count0<10 0\fi \the\count0
         }}
\newcommand{\timestamp}{{\small\vbox{\hbox{\tt\jobname.tex}
\hbox{\the\day/\the\month/\the\year, \clock}}}}

\newcommand{\CO}{\mathcal{O}}

\newcommand{\R}{\mathbb{R}}

\newcommand{\grad}{\vec{\nabla}}

\newcommand{\spa}{\ , \ \ }

\newcommand{\ds}{\displaystyle}

\newcommand{\tr}{\mathop{{\rm Tr}}}
\DeclareMathOperator{\sgn}{sgn}

\newcommand{\vecto}[2]{\left( \begin{array}{c} #1 \\ #2 \end{array}
\right) }
\newcommand{\matrto}[4]{\left( \begin{array}{cc} #1 & #2 \\
#3 & #4 \end{array} \right) }

\newcommand{\matrbig}[4]{\left( \begin{array}{cc} \displaystyle #1 & \displaystyle #2 \\[3mm]
\displaystyle #3 & \displaystyle #4 \end{array} \right) }
\newcommand{\matrtreb}[9]{\left( \begin{array}{ccc} \displaystyle #1
& \displaystyle #2 & \displaystyle #3 \\[2mm]
\displaystyle #4 & \displaystyle #5 & \displaystyle #6 \\[2mm]
\displaystyle #7 & \displaystyle #8 & \displaystyle #9 \end{array} \right) }

\newtheorem{definition}{Definition}[section]

\newtheorem{theorem}[definition]{Theorem}
\newtheorem{lemma}[definition]{Lemma}

\newcommand{\proof}{\noindent {\bf Proof:}\ }
\newcommand{\squ}{\noindent $\square$}

\numberwithin{equation}{section}

\begin{document}

\begin{titlepage}

\rightline{\vbox{\small\hbox{\tt hep-th/0508208} }}
\vskip 2.5cm

\centerline{\Large \bf On the Structure of} \vskip 0.15cm
\centerline{\Large \bf Stationary and Axisymmetric Metrics}

\vskip 1.6cm
\centerline{\bf Troels Harmark and Poul Olesen}
\vskip 0.5cm
\centerline{\sl The Niels Bohr Institute}
\centerline{\sl Blegdamsvej 17, 2100 Copenhagen \O, Denmark}

\vskip 0.5cm

\centerline{\small\tt harmark@nbi.dk, polesen@nbi.dk}

\vskip 1.6cm

\centerline{\bf Abstract} \vskip 0.2cm \noindent We study the
structure of stationary and axisymmetric metrics solving the
vacuum Einstein equations of General Relativity in four and higher
dimensions, building on recent work in hep-th/0408141. We write
the Einstein equations in a new form that naturally identifies the
sources for such metrics. The sources live in a one-dimensional
subspace and the entire metric is uniquely determined by them. We
study in detail the structure of stationary and axisymmetric
metrics in four dimensions, and consider as an example the sources
of the Kerr black hole.


\end{titlepage}

 \tableofcontents

\section{Introduction}

In recent years there have been a great deal of attention towards
research in black holes in higher-dimensional space-times. For
four-dimensional asymptotically flat space-times the Uniqueness
Theorems states that only one type of black holes exists for a
given set of asymptotic charges. In particular, for
four-dimensional gravity without matter, the Kerr black hole
\cite{Kerr:1963ud} is the unique black hole solution for a given
mass and angular momentum
\cite{Israel:1967wq,Carter:1971,Hawking:1972vc,Robinson:1975}.
Contrary to this, we have learned that for more than four
dimensions there are generically many available phases of black
holes.%
\footnote{Notice though the uniqueness theorems of
\cite{Gibbons:2002bh,Gibbons:2002av,Rogatko:2002bt} for
higher-dimensional static black holes in flat space-times.} For
Kaluza-Klein spaces, i.e. Minkowski-space times a circle, the
phase structure is very rich with interesting phase transitions
between different kinds of black holes, and in some cases even an
uncountable number of different phases are available
\cite{Kol:2004ww,Harmark:2005pp}. For five-dimensional
asymptotically flat space-times without matter we have in addition
to the Myers-Perry rotating black hole solution
\cite{Myers:1986un} also the recently discovered
rotating black ring solution of Emparan and Reall \cite{Emparan:2001wn}.%
\footnote{See also the recent review \cite{Horowitz:2005rs} and
references therein.}

To understand the phase structure of black holes in higher
dimensions it is important to search for new black hole solutions.
But the non-linearity of General Relativity makes it in general
very hard to find new solutions. However, a class of metrics for
which the Einstein equations simplify considerably are the
stationary and axisymmetric metrics. These are $D$-dimensional
metrics possessing $D-2$ mutually commuting Killing vector fields.
Among black hole metrics of this type are the Kerr rotating black
hole in four dimensions \cite{Kerr:1963ud}, the Myers-Perry
rotating black hole in five dimensions \cite{Myers:1986un} and the
rotating black ring \cite{Emparan:2001wn}. In
\cite{Harmark:2004rm} a canonical form for stationary and
axisymmetric metrics were found for which the Einstein equations
effectively reduce to a differential equation on an axisymmetric
$D-2$ by $D-2$ matrix
field living in three dimensional flat space.%
\footnote{Apart from a subclass of metrics that we will not treat
in this paper \cite{Harmark:2004rm}.} The results of
\cite{Harmark:2004rm} generalizes those of Papapetrou
\cite{Papapetrou:1953,Papapetrou:1966} for stationary and
axisymmetric metrics in four dimensions. Moreover, the results of
\cite{Harmark:2004rm} also generalizes the work of
\cite{Weyl:1917,Emparan:2001wk} on the Weyl-type metrics, which
are $D$-dimensional metrics with $D-2$ mutually commuting and
orthogonal Killing vector fields.
Ref.~\cite{Emparan:2001wk}%
\footnote{See Section 4.1 in \cite{Elvang:2004iz} for a brief
review of generalized Weyl solutions. See furthermore
\cite{Charmousis:2003wm} for work on extending the generalized
Weyl solutions of \cite{Emparan:2001wk} to space-times with a
cosmological constant.} is the higher-dimensional generalization
of the analysis of Weyl \cite{Weyl:1917} on static and
axisymmetric metrics in four dimensions.

In \cite{Harmark:2004rm} it is furthermore shown that any
stationary and axisymmetric solution have a certain structure of
its sources, called the rod-structure. The rod-structure is
connected to the three dimensional flat space that the reduced
Einstein equations can be seen to live in. In this
three-dimensional flat space, the sources for the metric consist
of thin rods lying along a certain axis of the space. To each of
these rods can be attached a direction in the $D-2$ dimensional
vector space that the Killing vector fields are spanning. The
analysis of stationary and axisymmetric solutions in terms of its
rod-structure generalizes the analysis of \cite{Emparan:2001wk}
for Weyl-type metrics, which again generalizes the analysis of
four-dimensional Weyl-type metrics (see for example
\cite{Stephani:2003} and references therein).

In this paper we find a new formulation of the reduced form of the
Einstein equations found in \cite{Harmark:2004rm}. In the new
formulation the sources for a given stationary and axisymmetric
solution are naturally identified. These sources consist of two
$D-2$ by $D-2$ matrix valued functions living on an axis in the
above-mentioned three-dimensional flat space that the reduced
version of the Einstein equations lives in. It is argued that
these sources uniquely determines the solution. One can therefore
say that the problem of finding a stationary and axisymmetric
solutions is reduced to a one-dimensional problem, i.e. the
problem of finding the appropriate sources for the solution.

We examine the general properties of the sources, which are
intimately connected to the rod-structure of a solution. As an
important part of this, we describe how to continue the sources
across rod endpoints. It was conjectured in \cite{Harmark:2004rm}
that a given rod-structure corresponds to a unique stationary and
axisymmetric metric. The idea being that the rod-structure for a
solution contains all information about the solution. However, we
do not find in this paper sufficient constraints to fix the
sources in terms of the rod-structure. We leave this as a problem
for future research.

For completeness we note here that there have been several recent
interesting developments in the study of stationary and
axisymmetric solutions in higher dimensions. In
\cite{Jones:2004pz,Jones:2005hj} a new tool to categorize
stationary and axisymmetric solutions by drawing the so-called
Weyl Card diagrams have been developed. Moreover, solution
generating techniques for five-dimensional solutions have been
explored in
\cite{Koikawa:2005ia,Mishima:2005id,Pomeransky:2005sj}. In
particular, in \cite{Pomeransky:2005sj} it is shown using
\cite{Harmark:2004rm} that Einstein equations are integrable for
stationary and axisymmetric metrics, and this is subsequently
employed in rederiving the Myers-Perry five-dimensional rotating
black hole.

In Section \ref{sec:rev} we review the results of
\cite{Harmark:2004rm} that we build on in this paper. In Section
\ref{sec:newform} we find the new formulation for the Einstein
equations, and we identify the sources for the stationary and
axisymmetric metrics. We furthermore examine the properties of
these sources. In Section \ref{sec:expA} we prove in the
four-dimensional case that stationary and axisymmetric solutions
are uniquely determined by their sources. We also comment on the
higher-dimensional cases. In Section \ref{sec:overlap} we find a
way to relate the behavior of the sources on each side of a rod
endpoint. We prove that we in principle can make a full
determination of the sources on one side of a rod endpoint by
knowing the sources on the other side of the endpoint. In Section
\ref{sec:asymp} we consider the asymptotic behavior of the sources
for asymptotically flat space-times. In Section \ref{sec:kerr} we
consider the example of the Kerr black hole, find its
matrix-valued potential and its sources, and we consider the
behavior of the sources near the endpoints of the rods. In Section
\ref{sec:concl} we present our conclusions.

\section{Review of stationary and axisymmetric solutions}
\label{sec:rev}

We review in the following some of the most important results of
\cite{Harmark:2004rm} for use in this paper. As mentioned in the
Introduction the analysis and results of \cite{Harmark:2004rm}
builds on and generalizes results of
Refs.~\cite{Weyl:1917,Papapetrou:1953,Papapetrou:1966,Emparan:2001wk}.
See the Introduction or \cite{Harmark:2004rm} for details on this.

\subsection{Canonical metric and equations of motion}

We consider in this section $D$-dimensional stationary and
axisymmetric solutions of the vacuum Einstein equations. These are
$D$-dimensional Ricci-flat manifolds with $D-2$ commuting linearly
independent Killing vector fields $V_{(i)}$, $i=1,...,D-2$. For
such solutions we can find coordinates $x^i$, $i=1,...,D-2$, along
with $r$ and $z$, so that
\begin{equation}
V_{(i)} = \frac{\partial}{\partial x^i} \ ,
\end{equation}
for $i=1,..,D-2$, and such that the metric is of the {\sl
canonical form} \cite{Harmark:2004rm}%
\footnote{Two further assumptions are required as premises in the
derivation (see \cite{Harmark:2004rm} for explanation): (i) The tensor
$V_{(1)}^{[\mu_1} V_{(2)}^{\mu_1} \cdots V_{(D-2)}^{\mu_{D-2}}
D^\nu V_{(i)}^{\rho]}$ vanishes at at least one point of the
manifold for a given $i=1,...,D-2$. (ii) $\det ( G_{ij} )$ is
non-constant on the manifold.}
\begin{equation}
\label{themet} ds^2 = \sum_{i,j=1}^{D-2} G_{ij} dx^i dx^j +
e^{2\nu} (dr^2+dz^2) \ ,
\end{equation}
with
\begin{equation}
\label{rdetG} r = \sqrt{ | \det ( G_{ij} ) | } \ .
\end{equation}
For a given $r$ and $z$ we can view $G_{ij}$ as a $D-2$ by $D-2$
real symmetric matrix with $G^{ij}$ as its inverse. With this, the
vacuum Einstein equations $R_{\mu \nu} = 0$ for the metric
\eqref{themet} with the constraint \eqref{rdetG} can be written as
\begin{eqnarray}
\label{Geq} & \ds G^{-1} \left( \partial_r^2 + \frac{1}{r}
\partial_r +
\partial_z^2 \right) G = (G^{-1} \partial_r G )^2 + (G^{-1}
\partial_z G )^2 \ ,
& \\ & \ds \label{nueqs}
\partial_r \nu = - \frac{1}{2r} + \frac{r}{8}
\tr \left( (G^{-1} \partial_r G )^2 - (G^{-1}\partial_z G )^2
\right) \, , \
\partial_z \nu = \frac{r}{4} \tr \left( G^{-1}
\partial_r G \, G^{-1} \partial_z G \right) \, .\ \ \ &
\end{eqnarray}
From this one sees that one can find stationary and axisymmetric
solutions of the vacuum Einstein equations by finding a solution
$G(r,z)$ of the equation \eqref{Geq}. Given $G(r,z)$ one can then
subsequently always find a solution $\nu(r,z)$ to \eqref{nueqs}
since these equations are integrable.

We can make a further formal rewriting of \eqref{Geq} by
recognizing that the derivatives respects the symmetries of a flat
three-dimensional Euclidean space with metric
\begin{equation}
\label{unmet} ds_3^2 = dr^2 + r^2 d\gamma^2 + dz^2 \ .
\end{equation}
Here $\gamma$ is an angular coordinate of period $2\pi$.%
\footnote{It is important to remark that $\gamma$ is not an actual
physical variable for the solution \eqref{themet}, but rather an
auxilirary coordinate that is useful for understanding the
structure of Eqs.~\eqref{Geq}.} Therefore, if we define $\grad$ to
be the gradiant in three-dimensional flat Euclidean space, we can
write \eqref{Geq} as
\begin{equation}
\label{formalG} G^{-1} \grad^2 G = ( G^{-1} \grad G)^2  \ .
\end{equation}
For use in the paper we elaborate here a bit more on the precise
meaning of the formula \eqref{formalG}. Using the auxilirary
coordinate $\gamma$ we can define the Cartesian coordinates
$\sigma_1$ and $\sigma_2$ as
\begin{equation}
\label{sigmadef} \sigma_1 = r \cos \gamma \spa \sigma_2 = r \sin
\gamma \ .
\end{equation}
Clearly, we have then that $(\sigma_1,\sigma_2,z)$ are Cartesian
coordinates for the flat three-dimensional Euclidean space defined
in \eqref{unmet} since the metric in these coordinates is
\begin{equation}
ds_3^2 = d\sigma_1^2 + d\sigma_2^2 + dz^2 \ .
\end{equation}
We write the Cartesian components of a three-dimensional vector as
$\vec{w} = (w_1,w_2,w_z)$, and the gradiant used in
\eqref{formalG} is then given by
\begin{equation}
\grad = \left( \partial_1, \partial_2 , \partial_z \right) \ ,
\end{equation}
where $\partial_1 = \partial / \partial \sigma_1$ and $\partial_2
=\partial / \partial \sigma_2$.

\subsection{Rod-structure of a solution}
\label{sec:rods}

If we consider the condition \eqref{rdetG} we see that
$\det ( G(0,z) ) = 0$. Therefore, the dimension of the kernel of
$G(0,z)$ must necessarily be greater than or equal to one
for any $z$, i.e. $\dim ( \ker ( G(0,z) )) \geq 1$.

In \cite{Harmark:2004rm} we learned that in order to avoid curvature
singularities it is a necessary condition on a solution
that $\dim ( \ker ( G(0,z) )) = 1$, except for isolated
values of $z$.
We therefore restrict ourselves to solutions where this applies.
Naming the isolated $z$-values as $a_1,...,a_N$, we see that
the $z$-axis splits up into the $N+1$ intervals $[-\infty,a_1]$,
$[a_1,a_2]$,...,$[a_{N-1},a_N]$, $[a_N,\infty]$.
For a given stationary and axisymmetric solution we thus have that
the $z$-axis at $r=0$ is divided into these intervals, called {\sl rods}.

Consider now a specific rod $[z_1,z_2]$.
From \cite{Harmark:2004rm} we know that we can
find a vector
\begin{equation}
v = v^i \frac{\partial}{\partial x^i}  \ ,
\end{equation}
such that
\begin{equation}
\label{Gveq}
G (0,z) v = 0 \ ,
\end{equation}
for $i=1,...,D-2$ and $z \in [z_1,z_2]$.
This vector $v$ is called the {\sl direction} of the rod $[z_1,z_2]$.

We note that if $G_{ij} v^i v^j / r^2$ is negative (positive) for
$r \rightarrow 0$ we say the rod $[z_1,z_2]$ is {\sl time-like}
({\sl space-like}). For a space-like rod $[z_1,z_2]$ we clearly
have a potential conical singularity for $r \rightarrow 0$ when $z
\in \, ] z_1,z_2 [$. However, if $\eta$ is a coordinate made as a
linear combination of $x^i$, $i=1,...,D-2$, with
\begin{equation}
\frac{\partial}{\partial \eta} = v = v^i \frac{\partial}{\partial
x^i} \ ,
\end{equation}
then we can cure the conical singularity at the rod by requiring
the coordinate $\eta$ to have the period
\begin{equation}
\label{deta}
\Delta \eta = 2\pi \lim_{r \rightarrow 0}
\sqrt{ \frac{r^2 e^{2\nu} }{G_{ij} v^i v^j} } \ ,
\end{equation}
for $z \in [z_1,z_2]$.

\section{New formulation and identification of sources}
\label{sec:newform}

In this section we rewrite the vacuum Einstein equations in a way
that gives a natural identification of the sources for a given
stationary and axisymmetric solution. The purpose of identifying
such sources is that one can hope to reduce the problem of finding
new solutions to the problem of finding the sources for new
solutions. We comment on the general philosophy behind the
introduction of these sources in the end of this section.

\subsection{New formulation}

We begin by defining the field $\vec{C} (r,z)$ by
\begin{equation}
\label{defC} \vec{C} = G^{-1} \grad G \ .
\end{equation}
Thus, $C_r = G^{-1} \partial_r G$ and $C_z = G^{-1} \partial_z G$.
With this definition, the equation \eqref{formalG} for $G(r,z)$ is
equivalent to the equation $\grad \cdot \vec{C} = 0$ for $r > 0$.
However, for $r=0$ we can have sources for $\vec{C}$. Therefore,
we require $\vec{C} (r,z)$ to obey the equation
\begin{equation}
\label{divC} \grad \cdot \vec{C} = 4\pi \delta^2 (\sigma) \rho (z)
\ ,
\end{equation}
i.e. that $r^{-1} \partial_r ( r C_r) + \partial_z C_z = 4\pi
\delta^2 (\sigma) \rho (z)$. Here $\delta^2 (\sigma) \equiv \delta
(\sigma_1 ) \delta (\sigma_2)$, where $\sigma_1$ and $\sigma_2$
are the Cartesian coordinates defined in \eqref{sigmadef}. Thus,
the delta-function $\delta^2 (\sigma)$ expresses that we have
sources for $\vec{C}(r,z)$ at $r=0$. The $D-2$ by $D-2$
matrix-valued function $\rho(z)$ in \eqref{divC} then
parameterizes the sources for $\vec{C}(r,z)$ at $r=0$.

Interestingly, the equation \eqref{divC} is now a linear first
order differential equation which is equivalent to the non-linear
second order equation \eqref{formalG}. However, the price of
introducing $\vec{C}$ is that $\vec{C}$ in addition should obey
the non-linear
equation%
\footnote{Note our conventions for the cross product is such that
the epsilon symbol has $\epsilon^{12z} =1$ which in $(r,\gamma,z)$
coordinates means that $\epsilon^{r\gamma z}=1$.}
\begin{equation}
\label{curlC} \grad \times \vec{C} + \vec{C} \times \vec{C} = 0 \
,
\end{equation}
i.e. that $\partial_r C_z + \partial_z C_r + [C_r,C_z] = 0$. The
equation \eqref{curlC} is found from $\grad \times (G^{-1} \grad
G) = \grad G^{-1} \times \grad G = - G^{-1} \grad G G^{-1} \times
\grad G = - ( G^{-1} \grad G ) \times ( G^{-1} \grad G )$. Note
that there are no source terms to the equation \eqref{curlC} so it
also holds for $r=0$. In conclusion, we have exchanged the
non-linear equation second order equation \eqref{formalG} with the
two first order equations \eqref{divC} and \eqref{curlC}.

We see from Eq.~\eqref{divC} that $C_{ij}(r,z)$ resembles an
electric field with $\rho_{ij}(z)$ being the charge density at
$r=0$. However, the $\vec{C} \times \vec{C}$ term of
Eq.~\eqref{curlC} introduces terms that mix the different
components in such a way that $\vec{C}(r,z)$ cannot be obtained
with linear superposition of fields.

The constraint \eqref{rdetG} on the determinant of $G$ is
equivalent to demanding
\begin{equation}
\label{trC} \tr ( \vec{C} ) = 2 \grad \log r \ ,
\end{equation}
i.e. that $\tr ( C_r ) = 2/r$ and $\tr( C_z) = 0$. Therefore, if
we take the trace of \eqref{divC} we get, using \eqref{trC} and
$\grad^2 \log r = 2\pi \delta^2 ( \sigma )$, the important
identity
\begin{equation}
\label{trrho} \tr ( \rho ) = 1 \ ,
\end{equation}
which holds for all $z$. Eq.~\eqref{trrho} is a generalization of
the requirement on the Generalized Weyl solutions of
\cite{Emparan:2001wk} that the total rod density is constant for
all $z$. As shown in \cite{Harmark:2004rm}, we get from this and
demanding absence of singularities for $r \rightarrow 0$ that
there is precisely one rod present for any given value of $z$
(except for isolated values of $z$).

Note that we can express the equations \eqref{nueqs} for
$\nu(r,z)$ in terms of $\vec{C}(r,z)$ as
\begin{equation}
\partial_r \nu = - \frac{1}{2r} + \frac{r}{8} \tr ( C_r^2 - C_z^2
) \spa
\partial_z \nu = \frac{r}{4} \tr ( C_r C_z ) \ .
\end{equation}
These equations ensure that we can find $\nu(r,z)$ from
$\vec{C}(r,z)$ alone.

It is interesting to consider how many different $G(r,z)$ that
corresponds to the same $\vec{C}(r,z)$. The answer is provided by
the following lemma:
\begin{lemma}
Let $G_1(r,z)$ and $G_2(r,z)$ be given such that $G_1^{-1} \grad
G_1 = G_2^{-1} \grad G_2$, with $G_1$ and $G_2$ being invertible
for $r \neq 0$. Then we have that $G_2 G_{1}^{-1}$ is a constant
matrix.\newline \proof Define $M = G_2 G_{1}^{-1}$. Then
\[
\begin{array}{rcl}
G_2^{-1} \grad G_2 &=& (M G_1)^{-1} \grad (M G_1) = G_1^{-1}
M^{-1} ( (\grad M) G_1 + M \grad G_1 ) \\[1mm] &=& G_1^{-1} M^{-1}
(\grad M) G_1 + G_1^{-1} \grad G_1 \ .
\end{array}
\]
We see from this that $\grad M = 0$ for $r \neq 0$. This means
that $M$ is a constant matrix. \squ
\end{lemma}

The above lemma means that given $\vec{C}(r,z)$ we can determine
the corresponding $G(r,z)$ up to a constant matrix, hence if we
for example require certain asymptotic boundary conditions on
$G(r,z)$ then we can determine $G(r,z)$ uniquely for a given
$\vec{C}(r,z)$.

\subsection{Defining a potential for $\vec{C}$}

We define the $D-2$ by $D-2$ matrix-valued function $A(r,z)$ by
\begin{equation}
\label{defA} C_r = - \frac{1}{r} \partial_z A \spa C_z =
\frac{1}{r} \partial_r A \ .
\end{equation}
This definition is meaningful for $r > 0$ since $r^{-1} \partial_r
(rC_r) + \partial_z C_z = 0$ implies the integrability condition
$\partial_r \partial_z A = \partial_z \partial_r A$. Hence, for a
given $\vec{C}$ we can find $A$ such that \eqref{defA} is
fulfilled for $r > 0$. The value of $A$ for $r=0$ is then found by
demanding continuity of $A$ at $r=0$.

Clearly, $A(r,z)$ is a potential for $\vec{C}(r,z)$. If we define
the field $\vec{A}(r,z)$ by $A_r(r,z)=A_z(r,z)=0$ and
$A_\gamma(r,z) = A(r,z)$ we can furthermore write that $\vec{C} =
\grad \times \vec{A}$. Therefore, the potential $A(r,z)$ can be
seen as a component of the vector potential $\vec{A}(r,z)$.%
\footnote{Note that in Cartesian coordinates
$\vec{A}=(A_1,A_2,A_z) = (A \partial_1 \gamma,A \partial_2
\gamma,0)$.} Note also that the definition of $A$ from $\vec{C}$
is ambiguous since if one uses the vector potential $\vec{A}' =
\vec{A} + \grad \alpha$ for a given $\alpha(r,\gamma,z)$ one gets
the same $\vec{C}$. However, demanding $A'_r=A'_z=0$ and that
$\partial_\gamma A' = 0$ we get that the definition \eqref{defA}
for $A(r,z)$ is only ambiguous up to an additive constant.

We can now use Gauss' law for $\vec{C}$ to derive an important
identity. Consider the cylindrical volume $V = \{ r \leq r_0 , z_1
\leq z \leq z_2 \}$. Then we have
\begin{equation}
\label{diffA} \begin{array}{l} \ds \frac{1}{2\pi} \int_{V} \grad
\cdot \vec{C} dV = \frac{1}{2\pi} \int_{\partial V} \vec{n} \cdot
\vec{C} dS
\\[3mm] \ds
= - \int_{r=0}^{r_0} \partial_r A |_{z=z_1} dr
 + \int_{r=0}^{r_0} \partial_r A |_{z=z_2} dr
 - \int_{z=z_1}^{z_2} \partial_z A |_{r=r_0} dz
 = A (0,z_1)  - A(0,z_2) \ .
\end{array}
\end{equation}
Using \eqref{divC} we then get%
\footnote{The identity \eqref{rhofromA} can also be derived using
the Cartesian coordinates \eqref{sigmadef}. Here one uses
$[\partial_1,\partial_2] \gamma = 2\pi \delta^2 (\sigma)$. Note
that just as for the Dirac monopole in electrodynamics one
encounters the Dirac string at $r=0$. The integral version of the
derivation \eqref{diffA} instead avoids this issue.}
\begin{equation}
\label{rhofromA} \rho(z) = - \frac{1}{2} \partial_z A (0,z) \ .
\end{equation}
Therefore, we get from \eqref{defA} that
\begin{equation}
\label{rhofromC} \rho(z) = \frac{1}{2} \lim_{r \rightarrow 0} r
C_r \ .
\end{equation}
We see then that $2\rho = \lim_{r\rightarrow 0} rG^{-1} \partial_r
G$ which means that one can compute $\rho$ from the metric
directly.

We have thus solved \eqref{divC} by introducing the potential
$A(r,z)$ via \eqref{defA} and demanding \eqref{rhofromA}. However,
we still have \eqref{curlC} which for $A(r,z)$ becomes
\begin{equation}
\label{Aeq} r \partial_r \left( \frac{1}{r} \partial_r A \right) +
\partial_z^2 A + \frac{1}{r} [ \partial_r A , \partial_z A ] = 0 \
.
\end{equation}
Thus, \eqref{formalG} for $G(r,z)$ have now been translated into
\eqref{Aeq} for $A(r,z)$. The constraint \eqref{rdetG}, which is
equivalent to \eqref{trC}, translates for the potential $A(r,z)$
to
\begin{equation}
\tr (A) = - 2z \ ,
\end{equation}
up to an additive constant.

\subsection{Behavior of a solution near $r=0$}
\label{sec:nearr0}

Equation \eqref{rhofromC} shows that $\rho(z)$ characterizes the
behavior of $C_r(r,z)$ for $r \rightarrow 0$. However, it will be
clear in the following that we in addition need to take into
account the behavior of $C_z(r,z)$ for $r\rightarrow 0$. We define
therefore the $D-2$ by $D-2$ matrix-valued function $\Lambda(z)$
by
\begin{equation}
\label{lambdef} \Lambda(z) = C_z (0,z) \ .
\end{equation}
We comment on the finiteness of $\Lambda(z)$ below. From
\eqref{trC} it is easily seen that
\begin{equation}
\label{trlamb} \tr ( \Lambda ) = 0 \ .
\end{equation}
Moreover, from considering the $r\rightarrow 0$ limit of
\eqref{curlC}, we get
\begin{equation}
\label{rholamb} \rho' = [\rho , \Lambda] \ .
\end{equation}
Therefore, if one knows $\Lambda(z)$ it is possible to find
$\rho(z)$, at least up to an additive constant matrix.

Consider now a rod $[z_1,z_2]$, as defined in
Section \ref{sec:rods}. Without loss of generality we consider the
rod $[z_1,z_2]$ to have direction $v=(1,0,...,0)$. We consider what
happens for general $v$ below.
Using the analysis of the metric near a
rod in \cite{Harmark:2004rm} we see then that
for $r \rightarrow 0$ and $z \in ]z_1,z_2[$
we have that $G_{11} = \CO(r^2)$, $G_{1i} = G_{i1} = \CO(r^2)$ and $G_{ij} = \CO(1)$ with $i,j=2,...,D-2$, while for the inverse we have
that $G^{11} = \CO(r^{-2})$,
$G^{1i} = G^{i1} = \CO(1)$ and $G^{ij} = \CO(1)$ with $i,j=2,...,D-2$.
Moreover, writing $G_{11} \simeq \pm e^{\lambda(z)} r^2 $, we have that
$G^{11} \simeq \pm e^{-\lambda(z)} r^{-2}$
for $r\rightarrow 0$ and $z \in ]z_1,z_2[$.
From \eqref{defC}
this gives then for $r \rightarrow 0$ and $z \in ]z_1,z_2[$
that $C_{r,11} \simeq 1/r$, $C_{r,1i} = \CO(r^{-1})$,
$C_{r,i1} = \CO(r)$ and $C_{r,ij} = \CO (r)$ with $i,j=2,...,D-2$,
while $C_{z,11} \simeq \lambda'(z)$, $C_{z,1i} = \CO(1)$,
$C_{z,i1} = \CO(r^2)$ and $C_{z,ij} = \CO (1)$ with $i,j=2,...,D-2$.

Using then \eqref{rhofromC} and \eqref{lambdef}
we conclude that both $\rho(z)$ and $\Lambda(z)$
are well-defined and smooth for $z \in ]z_1,z_2[$. Moreover,
for $\rho(z)$ we get
that $\rho_{11} = 1$ and that $\rho_{ij} = 0$ with $i=2,...,D-2$
and $j=1,...,D-2$ for $z \in ]z_1,z_2[$, while for
$\Lambda(z)$ we get that $\Lambda_{11} = \lambda'(z)$ and
$\Lambda_{i1} = 0$ with $i=2,...,D-2$ for $z \in ]z_1,z_2[$.
Note that we can infer from this that $\tr (\rho \Lambda ) = \lambda'$.
This is used below.

Consider now instead a given rod $[z_1,z_2]$ of direction $v$,
with $v$ being arbitrary. Clearly, we can find $\lambda(z)$ such
that $v^T G v = \pm \lambda(z) r^2 v^T v$. Using now the above
analysis along with Appendix \ref{sec:trans}, we can conclude for
$\rho(z)$ with $z \in ]z_1,z_2[$ that
\begin{equation}
\label{rhoprop} \rho v = v \spa \tr ( \rho ) = 1 \spa \rho^2 =
\rho  \ ,
\end{equation}
and that there exists linearly independent vectors
$w_1,...,w_{D-3}$ independent of $z$ such that
\begin{equation}
\rho^T w_i = 0 \ , \ i =1,...,D-3 \ .
\end{equation}
We see from \eqref{rhoprop} that $v$ is an eigenvector for $\rho$
with eigenvalue one. Concerning the properties involving
$\Lambda$, we have for $z \in ]z_1,z_2[$ that
\begin{equation}
\label{lambprop}
\Lambda v = \lambda' v \spa
\tr( \rho \Lambda ) = \lambda' \ .
\end{equation}
Thus, $v$ is also an eigenvector for $\Lambda(z)$.
If we consider $\nu(r,z)$ for $r\rightarrow 0$, we see from \eqref{nueqs} that
$\partial_r \nu \simeq -  ( 1 - \tr ( \rho^2 ) ) / (2r)$.
Using now \eqref{rhoprop} we see that $\tr (\rho^2 ) = 1$ and
hence that $\nu(0,z)$ is well-defined for $z\in ]z_1,z_2[$.
If we instead consider $\partial_z \nu$ for
$r \rightarrow 0$ we get from \eqref{nueqs}
and \eqref{lambprop} that
\begin{equation}
\label{partznu}
\partial_z \nu (0,z) = \frac{1}{2} \tr ( \rho \Lambda )
= \frac{1}{2} \lambda' (z) \ ,
\end{equation}
for $z\in ]z_1,z_2[$. Integrating this, we get $2\nu(0,z) =
\lambda (z) + \mbox{constant}$. This ensures that for space-like
rods we can remove the potential conical singularity by choosing a
suitable periodicity of the corresponding coordinate, while for
time-like rods it ensures the temperature of the corresponding
horizon is constant on the horizon. We see furthermore from
\eqref{partznu} that we can compute $\nu(0,z)$ for all $z$ by
knowing $\rho(z)$ and $\Lambda(z)$, and demanding continuity of
$\nu(0,z)$.

\subsubsection*{General philosophy}

The general philosophy of introducing the sources $\rho(z)$ and
$\Lambda(z)$ is that all information about the solution is stored
in these two matrix-valued function. This is shown explicitly in
the four-dimensional case in Section \ref{sec:expA}. Thus, we have
reduced the relevant physics to these two matrix-valued functions
living on a one-dimensional subspace. In particular, all the
physical quantities like the mass, angular momenta, temperature,
entropy, quadrupole moment, etc. can be obtained from $\rho(z)$
and $\Lambda(z)$ directly. We show this explicitly for the most
important asymptotic quantities in Section \ref{sec:asymp}.

The idea is moreover that the sources $\rho(z)$ and $\Lambda(z)$
should be uniquely determined by the rod-structure, i.e. by
specifying the division of the $z$-axis into separate intervals,
along with specifying the direction of these rods in the linear
vector space of Killing vector fields. This would be a
generalization of the solution generating technique for Weyl
solutions \cite{Weyl:1917,Emparan:2001wk}. In this paper we do not
go all the way to determine $\rho(z)$ and $\Lambda(z)$ from the
rod-structure, but in addition to the steps taken in this section
we take a further important step in this direction in Section
\ref{sec:overlap}.

\section{On uniqueness of $A(r,z)$ given $\rho(z)$ and $\Lambda(z)$}
\label{sec:expA}

The purpose of this section is to argue that stationary and
axisymmetric solutions of the vacuum Einstein equations are
uniquely determined by the sources $\rho(z)$ and $\Lambda(z)$
introduced in Section \ref{sec:newform}. We address this here by
arguing that the potential $A(r,z)$ introduced in Section
\ref{sec:newform} can be uniquely determined by $\rho(z)$ and
$\Lambda(z)$.

We prove in this section in detail that $A(r,z)$ is uniquely
determined by the two matrix-valued functions $\rho(z)$ and
$\Lambda(z)$ in four dimensions. This means that we can reduce the
question of finding a solution for $A(r,z)$ to finding the two
functions $\rho(z)$ and $\Lambda(z)$ that lives on a
one-dimensional subspace. We comment on the proof for arbitrary
dimensions in the end.

The general idea is to expand $A(r,z)$ around $r=0$. The expansion
is
\begin{equation}
A(r,z) = \sum_{n=0}^\infty A_n (z) r^{2n} \ .
\end{equation}
Since the behavior of $A(r,z)$ is singular around rod endpoints we
assume $z_1 < z < z_2$ where $[z_1,z_2]$ is one of the rods of the
solution. We get then from \eqref{Aeq}
\begin{eqnarray}
& \ds \label{A1A2} A_0'' = 2 [A_0' , A_1] \spa A_2   + \frac{1}{2}
[A_2,A_0'] = - \frac{1}{8} A_1'' - \frac{1}{4}[ A_1 , A_1' ] \ , &
\\[1mm] & \ds
\label{genAn} A_{n+1} + \frac{1}{2n} [ A_{n+1} , A_0' ] = -
\frac{1}{4n(n+1)} A_n'' - \sum_{k=1}^n \frac{n-l+1}{2n(n+1)} [
A_{n-k+1} , A_k' ] \ . &
\end{eqnarray}
Note that $A_0' = - 2 \rho$ and $A_1 = \frac{1}{2} \Lambda$.
Therefore, the basic idea in the following is to show that we can
find all $A_n$ for $n \geq 2$ from $A_0'$ and $A_1$.

We now restrict ourselves to four dimensions. We can assume
without loss of generality that we have a rod $[z_1,z_2]$ in the
direction
\begin{equation}
\vecto{1}{0} \ .
\end{equation}
Define
\begin{equation}
\alpha = \matrto{1}{0}{0}{0} \spa \beta = \matrto{0}{1}{0}{0} \spa
\gamma = \matrto{0}{0}{1}{0} \ .
\end{equation}
These three matrices have the properties
\begin{equation}
\begin{array}{c} \ds
\alpha^2 = \alpha \spa \beta^2 = 0 \spa \gamma^2 = 0 \spa \alpha
\beta = \beta \spa \beta \alpha = 0 \spa \alpha \gamma = 0 \spa
\gamma \alpha = \gamma \ ,
\\[1mm] \ds
\beta \gamma = \alpha \spa \gamma \beta = I - \alpha \spa
[\alpha,\beta] = \beta \spa [\alpha,\gamma] = - \gamma \spa
[\beta,\gamma] = 2\alpha - I \ .
\end{array}
\end{equation}
From the properties of $\rho$ of Section \ref{sec:nearr0} we see
that it should have the form $\rho(z) = \alpha + h(z) \beta$ where
$h(z)$ is a function. Moreover, the most general form of $\Lambda$
is $\Lambda(z) = \lambda'(z) ( 2 \alpha - I ) + ( h'(z) + 2 h(z)
\lambda'(z) ) \beta$ where $\lambda(z)$ is defined in Section
\ref{sec:nearr0}. Note that $A_0 = - 2 \alpha - 2 h \beta$. Since
$\tr ( A_n) = 0$ for $n \geq 1$ we can write
\begin{equation}
A_n = a_n \left( 2\alpha - I \right) + b_n \beta + c_n \gamma \ ,
\end{equation}
for $n \geq 1$ and $z_1 < z < z_2$. Note that $a_1 = \frac{1}{2}
q'$, $ b_1 = \frac{1}{2}h'+ hq' $ and $ c_1 = 0$. From
\eqref{genAn} we get then for $n \geq 1$
\begin{equation}
\begin{array}{c} \ds
4n(n+1) a_{n+1} + 4(n+1) h c_{n+1}  + a_n'' + 2 \sum_{k=1}^n
(n-k+1) ( b_{n-k+1} c_k' - c_{n-k+1} b_k' ) = 0 \ ,
\\[4mm] \ds
4(n+1)^2 b_{n+1} - 8(n+1) h a_{n+1} + b_n'' + 4 \sum_{k=1}^n
(n-k+1) ( a_{n-k+1} b_k' - b_{n-k+1} a_k' ) = 0 \ ,
\\[4mm] \ds
4(n-1)(n+1) c_{n+1} + c_n'' + 4 \sum_{k=1}^n (n-k+1) ( c_{n-k+1}
a_k' - a_{n-k+1} c_k' ) = 0 \ .
\end{array}
\end{equation}
For $n \geq 2$ we see that these three equations uniquely
determine $a_{n+1}$, $b_{n+1}$ and $c_{n+1}$ from $a_k$, $b_k$ and
$c_k$ for $k \leq n$. However, for $n=1$ these equations do not
constrain $c_2(z)$. Therefore $c_2(z)$ can be an arbitrary
function, according to these equations.

In order to fix $c_2(z)$ we should impose that we want $A(r,z)$ to
be a potential for $\vec{C}(r,z)$ such that $\vec{C} = G^{-1}
\grad G$ with $G^T = G$. Then we know from Appendix
\ref{sec:eom4D} that $h(z)$ and $\lambda'(z)$ can be chosen
freely. On the other hand, from the expansion of $G(r,z)$ in
Appendix \ref{sec:eom4D}, $h(z)$ and $\lambda'(z)$ are also the
only free functions. Therefore, $c_2(z)$ must be determined in
terms of $h$ and $\lambda'$. Indeed from \eqref{theA221} we get
\begin{equation}
\label{c2} c_{2}(z) = - \frac{1}{4} e^{2\lambda} h' \ .
\end{equation}
So, given $h(z)$ and $\lambda'(z)$, we must choose $c_2(z)$ as
\eqref{c2} in order for $A(r,z)$ to be a potential for which a
symmetric $G(r,z)$ exists. Said in another way, imposing
\eqref{Aeq} on $A(r,z)$ ensures the existence of a $G(r,z)$ such
that $\vec{C} = G^{-1} \grad G$, but imposing in addition that
$G^T = G$ means that we also have to fix $c_2(z)$.

We can formulate the above as a theorem:
\begin{theorem} \label{theo:uniqA}
Let a rod $[z_1,z_2]$ be given. Given $\rho(z)$ and $\Lambda(z)$
for $z_1 < z < z_2$ there exists precisely one $A(r,z)$ that
solves \eqref{Aeq} and that corresponds to a symmetric $G(r,z)$,
for $z_1 < z < z_2$.
\end{theorem}

To show that this theorem works in arbitrary dimensions one should
generalize the procedure above. One can of course without loss of
generalitity choose a rod of direction $(1,0,...,0)$ generalizing
the four dimensional case. The proof basically rest on having that
$A_{n+1}$ for a given $n$ is determined by Eq.~\eqref{genAn}. If
we take into account the general properties of $\rho$ from Section
\ref{sec:nearr0} we see that this is only problematic for $n=1$.
For $n=1$ we have that only the top entry of the first column of
the matrix $A_2 - [A_2 ,\rho]$ is non-zero, no matter what $A_2$
is. Therefore, given $\rho$ and $\Lambda$ we can choose $D-3$
additional arbitrary functions and still obey \eqref{Aeq},
generalizing the arbitrariness of $c_2(z)$ for four dimensions. We
expect these $D-3$ free functions to be determined by the $G^T =
G$ condition, just as in four dimensions.

\section{Smoothness conditions at rod endpoints}
\label{sec:overlap}

In this section we take a further step in understanding how to
obtain the sources $\rho(z)$ and $\Lambda(z)$ from specifying the
rod-structure for a solution. What we address in this section is
how to relate the sources on both sides of a rod endpoint. Using
this one can find the sources on one side of a rod endpoint from
knowing the sources on the other side of the rod endpoint.

We can address the question of how to connect the sources across
rod endpoints in a mathematical way as follows. We have seen that
$\rho(z)$ and $\Lambda(z)$ are smooth matrix valued functions away
from endpoints of rods. But, this is not true at the rod
endpoints. This is directly related to the fact that the potential
$A(r,z)$ is not smooth near rod endpoints. However, we want the
solution as a whole to be regular, thus we need a criteria for
when a solution is well-behaved near a rod endpoint. This is the
subject of this section.

Our basic requirement for smoothness of a solution is:
\begin{itemize}
\item Let $z_*$ be given. Define the coordinates
\begin{equation}
\label{pqdef} p = z - z_* \spa q = \sqrt{r^2 + (z-z_*)^2} \ .
\end{equation}
We then require of a solution that $A(p,q)$ should be smooth in a
neighborhood of $(p,q)=(0,0)$.
\end{itemize}
Away from the endpoints, this reduce to usual requirement of
smoothness of $A(r,z)$ near $r=0$, which already is contained in
the previous sections. However, if $z_*$ is an endpoint, i.e. if
we suppose that we are considering a solution with two different
rods $[z_1,z_*]$ and $[z_*,z_2]$ so that $z=z_*$ divides the two
rods, then the above requirement becomes non-trivial.

In the following we use the above smoothness requirement to
analyze the behavior of $\rho(z)$ and $\Lambda(z)$ across
endpoints. Let now $z_*$ be given, and consider the $(p,q)$
coordinates as defined in \eqref{pqdef}. We define
\begin{equation}
\label{Tmn} T^{(m,n)} = \partial^m_p \partial^n_q A
|_{(p,q)=(0,0)} \ ,
\end{equation}
as notation for the derivatives of $A(p,q)$, with
$T^{(0,0)}=A(0,0)$. Using that $2\rho = - \partial_p A |_{r=0} -
\sgn (p) \partial_q A |_{r=0}$ and $\Lambda = |p|^{-1}
\partial_q A |_{r=0}$, we find that the expansions of $\rho$
and $\Lambda$ around $z=z_*$ are given by
\begin{eqnarray}
& \ds \label{rhoT} \rho = \frac{1}{2} \sum_{k=0}^\infty
\frac{p^k}{k!} \sum_{l=0}^k \frac{k!}{l!(k-l)!} \left( - \sgn(p)^l
T^{(k-l+1,l)} - \sgn(p)^{l+1} T^{(k-l,l+1)} \right) \ , & \\[1mm] & \ds
\label{lambT} p \Lambda = \sum_{k=0}^\infty \frac{p^k}{k!}
\sum_{l=0}^k \frac{k!}{l!(k-l)!} \sgn(p)^{l+1} T^{(k-l,l+1)} \ . &
\end{eqnarray}
If we consider $T^{(0,1)}$ we see from \eqref{rhoT} and
\eqref{lambT} that
\begin{equation}
\label{ol1} \lim_{z \rightarrow z_*^-} \rho(z) - \lim_{z
\rightarrow z_*^+} \rho(z) = \lim_{z \rightarrow z_*^+} (z-z_*)
\Lambda (z) = - \lim_{z \rightarrow z_*^-} (z-z_*) \Lambda(z) \ .
\end{equation}
Considering instead $T^{(1,0)}$ and $T^{(2,0)} - T^{(0,2)}$, we
get
\begin{eqnarray}
\label{ol2} & \ds \lim_{z\rightarrow z_*+ } \left[ 2\rho(z) +
(z-z_*) \Lambda (z) \right] = \lim_{z\rightarrow z_*- } \left[
2\rho(z) + (z-z_*) \Lambda (z) \right] \ , &
\\[1mm] & \ds
\label{ol3} \lim_{z\rightarrow z_*+ } \left[ \rho'(z) + ( (z-z_*)
\Lambda (z))' \right] = \lim_{z\rightarrow z_*- } \left[ \rho'(z)
+ ( (z-z_*) \Lambda (z) )' \right] \ . &
\end{eqnarray}
Eqs.~\eqref{ol1}-\eqref{ol3} relates the behavior of $\rho(z)$ and
$\Lambda(z)$ for $z \rightarrow z_*^+$ to their behavior for $z
\rightarrow z_*^-$. This means that knowing $\rho(z)$ and
$\Lambda(z)$ for $z
> z_*$ we can get information on $\rho(z)$ and $\Lambda(z)$ for $z
< z_*$. The question is now whether one can fully determine
$\rho(z)$ and $\Lambda(z)$ for $z < z_*$. This is in fact possible
though one needs to use the extra constraints on $T^{(p,q)}$
coming from expanding
\begin{equation}
\label{Aeqpq}
\partial_p^2 A + \partial_q^2 A + \frac{2p}{q} \partial_p
\partial_q A = \frac{1}{q} [ \partial_p A , \partial_q A ] \ ,
\end{equation}
around $(p,q)=(0,0)$. Eq.~\eqref{Aeqpq} is Eq.~\eqref{Aeq} in
$(p,q)$ coordinates. The first few terms of this expansion give
\begin{equation}
\begin{array}{c}
\ds [T^{(1,0)},T^{(0,1)}] = 0 \spa 2 T^{(1,1)} =
[T^{(2,0)},T^{(0,1)}] +[T^{(1,0)},T^{(1,1)}] \ ,
\\[1mm]
T^{(2,0)} + T^{(0,2)} = [T^{(1,0)},T^{(1,2)}]
+[T^{(1,1)},T^{(0,1)}] \ .
\end{array}
\end{equation}

That we can connect $\rho(z)$ and $\Lambda(z)$ across a rod
endpoint is formulated in the following theorem:
\begin{theorem}
Let two rods $[z_1,z_*]$ and $[z_*,z_2]$ be given. Assume that we
know $\rho(z)$ and $\Lambda(z)$ for $z_1 < z < z_*$. Then we can
determine $\rho(z)$ and $\Lambda(z)$ for $z_* < z < z_2$ uniquely
by the smoothness condition on $A(p,q)$ formulated above with
$(p,q)$ given in \eqref{pqdef}.
\end{theorem}
We can prove this as follows. Since we know $\rho(z)$ and
$\Lambda(z)$ for $z_1 < z < z_*$ we can determine $A(r,z)$ for
$z_1 < z < z_*$ using
Theorem \ref{theo:uniqA}.%
\footnote{We assume here the validity of Theorem \ref{theo:uniqA}
for all dimensions, although we strictly speaking only have proven
Theorem \ref{theo:uniqA} in detail in four dimensions.} Employing
the coordinate transformation \eqref{pqdef} we can then find
$A(p,q)$ for $p < 0$ and $q$ in a certain range. Since $A(p,q)$ is
assumed to be smooth in $(p,q)=(0,0)$ we see that all the
derivatives of $A(p,q)$ at $(p,q)=(0,0)$ are the same whether one
approaches from $p < 0$ or $p > 0$. Thus, we can determine
$T^{(m,n)}$ as defined by \eqref{Tmn} from our knowledge of
$\rho(z)$ and $\Lambda(z)$ for $z_1 < z < z_*$. Using now
\eqref{rhoT} and \eqref{lambT} we can find the values of the
derivatives of $\rho(z)$ and $(z-z_*) \Lambda(z)$ as $z
\rightarrow z_*^+$ to all orders. This then uniquely determines
$\rho(z)$ and $\Lambda(z)$ for $z_* < z < z_2$.

\section{Asymptotically flat space-times}
\label{sec:asymp}

In this section we consider solutions that asymptote to four- and
five-dimensional Minkowski spaces. We work in units where Newtons
constant is set to one.

An important purpose of this section is to show that one can find
the asymptotic quantities like mass and angular momenta directly
from the sources $\rho(z)$ and $\Lambda(z)$. This is an
alternative way of viewing the statement that all the information
about a solution is contained in the sources $\rho(z)$ and
$\Lambda(z)$.

\subsubsection*{Four-dimensional asymptotically flat space-times}

From \cite{Harmark:2004rm} we know the asymptotic behavior of four
dimensional solutions asymptoting to Minkowski space. Using this,
we get the following asymptotic behavior for $A(r,z)$
\begin{equation}
A_{11} = - \frac{2Mz}{\sqrt{r^2+z^2}} \spa A_{12} =
\frac{2Jz(3r^2+2z^2)}{(r^2+z^2)^{3/2}} \spa A_{21} = -
\frac{2Jz}{(r^2+z^2)^{3/2}} \ ,
\end{equation}
in the asymptotic region $\sqrt{r^2+z^2} \rightarrow \infty$ with
$z/\sqrt{r^2+z^2}$ finite. Here $M$ is the mass and $J$ is the
angular momentum. Combining this with Eq.~\eqref{diffA} and
Eq.~\eqref{divC} we get
\begin{equation}
\int_{-\infty}^\infty dz \rho_{11} (z) = 2M \spa
\int_{-\infty}^\infty dz \rho_{12} (z) = -4J \spa
\int_{-\infty}^\infty dz \rho_{21} (z) = 0 \ .
\end{equation}
Using Eq.~\eqref{rhofromC} we get that $\rho(z)$ and $\Lambda(z)$
asymptotically behave as
\begin{equation}
\rho(z) = \matrbig{0}{\ 0}{-\frac{2J}{|z|^3}}{\ 1} \spa \Lambda(z)
= \sgn(z)
\matrbig{\frac{2M}{z^2}}{0}{\frac{6J}{z^4}}{-\frac{2M}{z^2}} \ ,
\end{equation}
for $z \rightarrow \pm \infty$. Note that the entries with zeroes
are zero to all orders in $1/z$ as can be seen from the properties
of $\rho$ and $\Lambda$ found in Section \ref{sec:nearr0}.

\subsubsection*{Five-dimensional asymptotically flat space-times}

We know from \cite{Harmark:2004rm} the asymptotics of solutions
that asymptote to five-dimensional Minkowski space. Using these
results we find the following asymptotic behavior of $A(r,z)$
\begin{equation}
\label{Aas5D}
\begin{array}{c} \ds
A_{11} = - \frac{4M}{3\pi} \frac{z}{\sqrt{r^2+z^2}} \spa A_{21} =
-\frac{J_1}{\pi} \frac{\sqrt{r^2+z^2}+z}{r^2+z^2} \spa A_{31} =
\frac{J_2}{\pi} \frac{\sqrt{r^2+z^2}-z}{r^2+z^2} \ ,
 \\[4mm] \ds
  A_{12} =
\frac{2J_1}{\pi} \frac{2z\sqrt{r^2+z^2}+r^2}{r^2+z^2} \spa A_{22}
= -(\sqrt{r^2+z^2}+z) + \frac{2(M+\eta)}{3\pi}
\frac{z}{\sqrt{r^2+z^2}} \ ,
 \\[4mm] \ds
A_{13} = \frac{2J_2}{\pi} \frac{2z\sqrt{r^2+z^2}-r^2}{r^2+z^2}
\spa A_{33} = \sqrt{r^2+z^2}-z + \frac{2(M-\eta)}{3\pi}
\frac{z}{\sqrt{r^2+z^2}} \ ,
 \\[4mm] \ds
A_{23} = \zeta \frac{r^2+2z^2+2z\sqrt{r^2+z^2}}{(r^2+z^2)^{3/2}}
\spa A_{32} = -\zeta
\frac{r^2+2z^2-2z\sqrt{r^2+z^2}}{(r^2+z^2)^{3/2}} \ ,
\end{array}
\end{equation}
in the asymptotic region $\sqrt{r^2+z^2} \rightarrow \infty$ with
$z/\sqrt{r^2+z^2}$ finite. Here $M$ is the mass and $J_1$ and
$J_2$ are the angular momenta. See \cite{Harmark:2004rm} for
comments on $\eta$ and $\zeta$. Using now Eq.~\eqref{Aas5D}
together with Eq.~\eqref{diffA} and Eq.~\eqref{divC} we get
\begin{equation}
\begin{array}{c} \ds
\int_{-\infty}^\infty dz \rho_{11} (z) = \frac{4M}{3\pi} \spa
\int_{-\infty}^\infty dz \rho_{12} (z) = -\frac{4J_1}{\pi} \spa
\int_{-\infty}^\infty dz \rho_{13} (z) = -\frac{4J_2}{\pi} \ ,
\\[4mm] \ds
\int_{-\infty}^\infty dz \rho_{21} (z) =\int_{-\infty}^\infty dz
\rho_{31} (z) =\int_{-\infty}^\infty dz \rho_{23} (z)
=\int_{-\infty}^\infty dz \rho_{32} (z) =0 \ .
\end{array}
\end{equation}
Finally, we see from Eq.~\eqref{Aas5D} using Eq.~\eqref{rhofromC}
that to leading order
\begin{equation}
\rho(z) = \matrtreb{0}{0}{0}{-\frac{J_1}{\pi
z^2}}{1}{\frac{2\zeta}{z^2}}{0}{0}{0} \spa \Lambda(z) =
\matrtreb{0}{0}{0}{0}{-\frac{1}{z}}{0}{0}{0}{\frac{1}{z}} +
\matrtreb{\frac{4M}{3\pi z^2}}{0}{-\frac{8J_2}{\pi
z^2}}{\frac{3J_1}{\pi z^3}}{ - \frac{2(M+\eta)}{3\pi
z^2}}{-\frac{8\zeta}{z^3}}{\frac{J_2}{\pi z^3}}{0}{-
\frac{2(M-\eta)}{3\pi z^2}}  \ ,
\end{equation}
for $z \rightarrow \infty$ and
\begin{equation}
\rho(z) = \matrtreb{0}{0}{0}{0}{0}{0}{-\frac{J_2}{\pi
z^2}}{\frac{2\zeta}{z^2}}{1} \spa \Lambda(z) =
\matrtreb{0}{0}{0}{0}{\frac{1}{z}}{0}{0}{0}{-\frac{1}{z}} +
\matrtreb{-\frac{4M}{3\pi z^2}}{\frac{8J_1}{\pi
z^2}}{0}{\frac{J_1}{\pi z^3}}{\frac{2(M+\eta)}{3\pi
z^2}}{0}{\frac{3J_2}{\pi
z^3}}{-\frac{8\zeta}{z^3}}{\frac{2(M-\eta)}{3\pi z^2}} \ ,
\end{equation}
for $z \rightarrow -\infty$.

\section{Example: The Kerr black hole}
\label{sec:kerr}

To illustrate the methods of this paper to analyze stationary and
axisymmetric solutions, we consider in detail the Kerr black hole
\cite{Kerr:1963ud}.

The Kerr metric is formulated most easily in the prolate spherical
coordinates $(x,y)$ defined as
\begin{equation}
r = \alpha \sqrt{(x^2-1)(1-y^2)} \spa z = \alpha x y \spa x \geq 1
\spa -1 \leq y \leq 1 \ ,
\end{equation}
where $\alpha > 0 $ is a number. The metric is then given as
\begin{equation}
G=\matrto{-X}{-X\tilde{A}~}{-X\tilde{A}}{X^{-1}(r^2-\tilde{A}^2X^2)}
\ ,
\end{equation}
\begin{equation}
X = \frac{x^2 \cos^2 \lambda + y^2 \sin^2 \lambda -1}{\left( 1 + x
\cos \lambda \right)^2 +  y^2 \sin^2 \lambda} \spa \tilde{A} =
\frac{2\alpha \tan \lambda (1-y^2)(1 + x \cos \lambda )}{x^2
\cos^2 \lambda + y^2 \sin^2 \lambda -1} \ .
\end{equation}
We now find the potential $A$ for the Kerr metric. For this we
record that in prolate spherical coordinates $\vec{C} = \grad
\times \vec{A}$ becomes
\begin{equation}
C_x = - \frac{1}{\alpha(x^2-1)} \partial_y A \spa C_y =
\frac{1}{\alpha(1-y^2)} \partial_x A \ .
\end{equation}
Using this we get the following potential for the Kerr black hole
\begin{equation}
\label{AKerr}
\begin{array}{c} \ds
A_{11} = -\frac{2\alpha y(( 1+x\cos\lambda )^2+\sin^2\lambda
)}{\cos\lambda ((1+x\cos\lambda )^2+ y^2\sin^2\lambda) } \spa
A_{21}=-\frac{2y\sin\lambda }{(1+x\cos\lambda )^2+y^2\sin^2\lambda
} \ ,
\\[4mm] \ds
A_{12}=6\alpha^2~y\left(1-\frac{y^2}{3}\right)\frac{\sin\lambda}
{\cos^2\lambda}+\frac{2\alpha^2~y(1-y^2)^2\sin\lambda\tan^2\lambda}{(1+x\cos
\lambda )^2+y^2\sin^2\lambda } \ .
\end{array}
\end{equation}
From \cite{Harmark:2004rm} we have that the rod-structure of the
Kerr black hole consist of a space-like rod $[-\infty,-\alpha]$ in
the direction $(0,1)$, a time-like rod $[-\alpha,\alpha]$ in the
direction $(1,\Omega)$ and a space-like rod $[\alpha,\infty]$ in
the direction $(0,1)$. Here $\Omega$ is given as
\begin{equation}
\Omega = \frac{\sin \lambda \cos \lambda}{2\alpha ( 1 + \cos
\lambda)} \ .
\end{equation}
We now find the source distributions $\rho(z)$ and $\Lambda(z)$
for the Kerr metric. For $z < -\alpha$ we have
\begin{equation}
\label{rholamb1}
\begin{array}{c} \ds
\rho = \matrto{0}{0}{h}{1} \spa h(z) = - \frac{2\sin \lambda \cos
\lambda \left( 1- \frac{z}{\alpha} \cos \lambda \right)}{\alpha
\left( \left( 1- \frac{z}{\alpha} \cos \lambda \right)^2 + \sin^2
\lambda \right)^2} \ ,
\\[7mm] \ds
\Lambda = \matrto{-\lambda'}{0}{h'+2h\lambda'~}{\lambda'} \spa
\lambda'(z) = \frac{2\left[ \left(1 - \frac{z}{\alpha} \cos
\lambda \right)^2 - \sin^2 \lambda\right] }{\alpha \cos \lambda
\left( 1- \frac{z^2}{\alpha^2} \right) \left[ \left( 1 -
\frac{z}{\alpha} \cos \lambda \right)^2  + \sin^2 \lambda \right]
} \ .
\end{array}
\end{equation}
This corresponds to the $[-\infty,-\alpha]$ rod. For $z > \alpha$
we have
\begin{equation}
\label{rholamb3}
\begin{array}{c} \ds
\rho = \matrto{0}{0}{h}{1} \spa h(z) = - \frac{2\sin \lambda \cos
\lambda \left( 1+ \frac{z}{\alpha} \cos \lambda \right)}{\alpha
\left( \left( 1+ \frac{z}{\alpha} \cos \lambda \right)^2 + \sin^2
\lambda \right)^2} \ ,
\\[7mm] \ds
\Lambda = \matrto{-\lambda'}{0}{h'+2h\lambda'~}{\lambda'} \spa
\lambda'(z) = \frac{2\left[ \left(1 + \frac{z}{\alpha} \cos
\lambda \right)^2 - \sin^2 \lambda\right] }{\alpha \cos \lambda
\left( 1- \frac{z^2}{\alpha^2} \right) \left[ \left( 1 +
\frac{z}{\alpha} \cos \lambda \right)^2  + \sin^2 \lambda \right]
} \ .
\end{array}
\end{equation}
This corresponds to the $[\alpha,\infty]$ rod. Finally for
$-\alpha < z < \alpha$ we have
\begin{equation}
\label{rholamb2}
\begin{array}{c} \ds
\rho = \matrto{1-\Omega h}{h}{\Omega-\Omega^2 h~}{\Omega h} \spa
h(z) = \frac{1}{\Omega} - \frac{\sin \lambda \left( (1+\cos
\lambda)^2 - \frac{z^2}{\alpha^2} \sin^2 \lambda \right)}{\Omega^2
\alpha \left( (1+\cos \lambda)^2 + \frac{z^2}{\alpha^2} \sin^2
\lambda \right)^2} \ ,
\\[7mm] \ds
\Lambda = \matrto{-\Omega h' + (1-2\Omega
h)\lambda'}{h'+2h\lambda'}{-\Omega^2 h' + 2\Omega(1-\Omega
h)\lambda'~}{\Omega h' - (1-2\Omega h)\lambda'}  \ ,
\\[5mm] \ds
\lambda'(z) = \frac{4(1+\cos \lambda) z}{\alpha^2
\left(1-\frac{z^2}{\alpha^2}\right)\left[ (1+\cos \lambda)^2  +
 \frac{z^2}{\alpha^2} \sin^2 \lambda  \right]} \ .
\end{array}
\end{equation}
corresponding to the $[-\alpha,\alpha]$ rod.

One can now easily check that $\rho(z)$ and $\Lambda(z)$ given
above for the Kerr black hole obeys the properties found in
Section \ref{sec:nearr0}. It is also interesting to consider the
smoothness conditions of Section \ref{sec:overlap} on the
endpoints of the rods. Consider the endpoint $z=\alpha$. Using
\eqref{rholamb2} we compute
\begin{equation}
\label{leftt}
\begin{array}{c} \ds \lim_{z\rightarrow \alpha^-} \rho = \matrto{1}{0}{\Omega}{0}
\spa \lim_{z\rightarrow \alpha^-} (z-\alpha) \Lambda =
\matrto{-1}{0}{-2\Omega}{1} \ ,
 \\[6mm] \ds
\lim_{z\rightarrow \alpha^-} ( \rho' + ((z-\alpha) \Lambda)' )  =
\frac{1-2\cos \lambda}{2\alpha} \matrto{1}{0}{2\Omega}{-1} \ .
\end{array}
\end{equation}
Using \eqref{rholamb3} we compute
\begin{equation}
\label{rightt}
\begin{array}{c} \ds \lim_{z\rightarrow \alpha^+} \rho = \matrto{0}{0}{-\Omega}{1}
\spa \lim_{z\rightarrow \alpha^+} (z-\alpha) \Lambda =
\matrto{1}{0}{2\Omega}{-1} \ ,
 \\[6mm] \ds
\lim_{z\rightarrow \alpha^+} ( \rho' + ((z-\alpha) \Lambda)' )  =
\frac{1-2\cos \lambda}{2\alpha} \matrto{1}{0}{2\Omega}{-1} \ .
\end{array}
\end{equation}
From \eqref{leftt} and \eqref{rightt} one can now easily check
explicitly that \eqref{ol1}-\eqref{ol3} are obeyed. Similarly one
can check the smoothness conditions involving higher derivatives
of $\rho$ and $\Lambda$.

\section{Conclusions}
\label{sec:concl}

In this paper we have made a new formulation of the Einstein
equations for stationary and axisymmetric metrics. This was done
in Section \ref{sec:newform} by introducing the field $\vec{C} =
G^{-1} \grad G$ and its vector potential $\vec{A}$ given by
$\vec{C} = \grad \times \vec{A}$. This enabled us to naturally
identify the sources $\rho(z)$ and $\Lambda(z)$ for a given
solution. As argued in Section \ref{sec:expA} a solution should be
completely determined once $\rho(z)$ and $\Lambda(z)$ are given.
Hence, we have reduced the problem of finding stationary and
axisymmetric solution to the problem of specifying the sources
$\rho(z)$ and $\Lambda(z)$. This means that we have reduced the
problem of finding solutions to a one-dimensional problem, since
$\rho(z)$ and $\Lambda(z)$ are matrix-valued functions of just one
variable.

That all information about a solution is contained in the sources
$\rho(z)$ and $\Lambda(z)$ is also confirmed by the fact that one
can extract all physical quantities from these. In particular, in
Section \ref{sec:asymp} we consider how to find the most important
asymptotically measurable quantities for four- and
five-dimensional asymptotically flat solutions.

The remaining problem is to find an effective method to determine
$\rho(z)$ and $\Lambda(z)$ solely from information about the
rod-structure of the solution. A step towards this is provided in
Section \ref{sec:newform} in which we analyze the constraints on
$\rho(z)$ and $\Lambda(z)$ coming from the $r\rightarrow 0$ limit
of the Einstein equations, and from imposing a particular
rod-structure. These constraints significantly reduce the freedom
of choice for $\rho(z)$ and $\Lambda(z)$, in particular for the
four-dimensional case the left-over freedom is in terms of a
single function of $z$. A further step towards restricting
$\rho(z)$ and $\Lambda(z)$ from the imposed rod-structure is taken
in Section \ref{sec:overlap} where we analyze the conditions on
$\rho(z)$ and $\Lambda(z)$ when crossing a rod endpoint. However,
we do not find a sufficient amount of restrictions on $\rho(z)$
and $\Lambda(z)$ to determine them completely from the given
rod-structure. We leave therefore this as an open problem for
future research.

In Section \ref{sec:kerr} we applied our considerations to the
example of the Kerr black hole. We found the potential $A(r,z)$
for the Kerr black hole and from that the sources $\rho(z)$ and
$\Lambda(z)$. Moreover, we checked successfully that the
conditions of Section \ref{sec:overlap} on the specific sources
when crossing rod endpoints are obeyed.

In conclusion, we have accomplished several steps towards the goal
of this paper, which is to find a general method of finding
stationary and axisymmetric metrics given a specific
rod-structure. However, we are still lacking a complete
determination of the sources. Despite this, it is clear from the
uniqueness theorems in four dimensions that one for example should
be able to determine completely the sources of the Kerr black
hole. On the other hand, we do not now of any direct construction
of the Kerr black hole. Indeed, if one determines $\rho(z)$ and
$\Lambda(z)$ directly from the rod-structure of the Kerr black
hole it will be the first example of such a direct construction.
One can therefore say that by devising a method to find the
sources solely from the rod-structure one would be obtaining a
generalization of the uniqueness theorems in four dimensions to
stationary and axisymmetric solutions in higher dimensions.

Clearly, the problem of finding the sources from the rod-structure
is interesting also with respect to the search for new black hole
solutions, in particular the five-dimensional asymptotically flat
black ring with two angular momenta, generalizing
\cite{Emparan:2001wn}, and non-zero angular momenta solutions with
Kaluza-Klein bubbles and event horizons \cite{Elvang:2004iz}.

\section*{Acknowledgments}

We thank Micha Berkooz for useful discussions. Work partially
supported by the European Community's Human Potential Programme
under contract MRTN-CT-2004-005104 `Constituents, fundamental
forces and symmetries of the universe'.

\begin{appendix}

\section{Rigid transformations}
\label{sec:trans}

Consider a coordinate transformation
\begin{equation}
\label{transM} \tilde{x}^i = \sum_j M_{ij} x^j \ .
\end{equation}
Here $M$ can be any (constant) invertible matrix. We see then that
\begin{equation}
\label{traGCA} \tilde{G} = M^{-T} G M^{-1} \spa \vec{\tilde{C}} =
M \vec{C} M^{-1} \spa \tilde{A} = M A M^{-1} \ .
\end{equation}
Moreover, we have
\begin{equation}
\label{haha} v^i = \sum_j M_{ij} v^j \spa \frac{\partial}{\partial
\tilde{x}^i} = \sum_j (M^{-T})_{ij} \frac{\partial}{\partial x^j}
\ .
\end{equation}
With a slight abuse of notation we write the first equation of
\eqref{haha} as
\begin{equation}
\tilde{v} = M v \ ,
\end{equation}
i.e. in this equation we read $v$ and $\tilde{v}$ as vectors in
$\R^{D-2}$ instead of vector fields in the $(D-2)$-dimensional
linear space spanned by the Killing vectors (clearly $v^i
\partial / \partial x^i$ is not affected by the transformation
\eqref{transM}). Thus, the point of the above is that by a
transformation \eqref{transM} we can transform a given non-zero
vector $v$ into any non-zero vector that we want. We just need to
take into account \eqref{traGCA}.

\section{Analysis of EOMs in four dimensions}
\label{sec:eom4D}

In this appendix we consider the expansion of the EOMs \eqref{Geq}
for $G$ around $r=0$ in four dimensions. In doing this, we show
that a solution of the EOMs is completely determined by two
functions. We use this in Section \ref{sec:expA}.

We parameterize $G$ as
\begin{equation}
G=\matrto{-X}{-X\tilde{A}~}{-X\tilde{A}}{X^{-1}(r^2-\tilde{A}^2X^2)}
\ .
\end{equation}
The EOMs \eqref{Geq} in terms of $X$ and $\tilde{A}$ are
\begin{eqnarray}
& \ds \label{eqXA1} X \left( \partial_r^2 + \frac{1}{r} \partial_r
+
\partial_z^2 \right) X = (\partial_r X)^2 + (\partial_z X)^2 -
\frac{X^4}{r^2} \left((\partial_r \tilde{A})^2 + (\partial_z
\tilde{A})^2\right)\ , & \\[1mm] & \ds \label{eqXA2}
\partial_r \left( \frac{X^2}{r} \partial_r \tilde{A} \right)
+ \partial_z \left( \frac{X^2}{r} \partial_z \tilde{A} \right) =
0\ . &
\end{eqnarray}
We consider now the behavior of a solution for $r$ small (i.e.
near $r=0$) and for $z$ away from the endpoints of any rod. Then
we can expand $X$ and $\tilde{A}$ as
\begin{equation}
X(r,z) = \sum_{n=0}^\infty f_n (z) r^{2n} \spa \tilde{A}(r,z) =
\sum_{n=0}^\infty g_n (z) r^{2n}\ .
\end{equation}
Considering then the leading part of \eqref{eqXA1} we get that
either $f_0(z)$ is zero, or $g_0(z)$ is constant.

We consider first the case in which $g_0(z)$ is constant. We write
\begin{equation}
f_0 (z) = e^{-\lambda(z)} \spa g_0(z) = - a \spa g_1(z) = - h(z)
e^{2\lambda(z)}\ .
\end{equation}
The first few terms in $X$ and $\tilde{A}$ then takes the form
\begin{equation}
\begin{array}{c} \ds
X = e^{-\lambda(z)} + r^2 f_1(z) + r^4 f_2(z) + \CO (r^6)\ ,
\\[2mm] \ds
\tilde{A} = -a - r^2 h(z) e^{2\lambda(z)} + r^4 g_2(z) + r^6
g_3(z) + \CO(r^8)\ .
\end{array}
\end{equation}
One can now find $f_1(z)$ and $g_2(z)$ from the EOMs
\eqref{eqXA1}-\eqref{eqXA2}:
\begin{equation}
f_1 = \frac{1}{4} e^{-\lambda} \lambda'' - e^{\lambda} h^2 \spa
g_2 = \frac{1}{8} e^{2\lambda} ( h'' + 4 h \lambda'' + 2 h'
\lambda' ) - e^{4\lambda} h^3\ .
\end{equation}
By going order by order one can similar find $f_2(z)$ and
$g_3(z)$, and then $f_3(z)$ and $g_4(z)$, and so on. Therefore,
once $h(z)$ and $\lambda(z)$ are given, $X(r,z)$ and
$\tilde{A}(r,z)$ are uniquely determined. We can now find $\rho$
and $\Lambda$:
\begin{equation}
\label{resrl} \rho = \matrto{a h}{\ a(1-a h)}{h}{1-ah} \spa
\Lambda = \matrto{a(h'+2h\lambda')-\lambda'}{\
-a^2(h'+2h\lambda')+2a\lambda'}{h'+2h\lambda'}{-a(h'+2h\lambda')+\lambda'}\
.
\end{equation}
One sees from this that if one knows both $\rho$ and $\Lambda$
then $X(r,z)$ and $\tilde{A}(r,z)$ are uniquely determined.
(Notice though the constant part of $\lambda(z)$. However, that is
not essential since that can be scaled to whatever one wants).

We consider now the case for which $f_0(z)=0$. We write
\begin{equation}
f_0(z) = 0 \spa f_1(z) = e^{\lambda(z)} \spa g_0(z) = h(z)\ .
\end{equation}
We then get
\begin{equation}
\label{resr2} \rho = \matrto{1}{h}{0}{0} \spa \Lambda =
\matrto{\lambda'}{h'+2h\lambda'}{0}{-\lambda'}\ .
\end{equation}
Writing
\begin{equation}
C_z(r,z) = \Lambda (z) + 4 r^2 A_2(z) + \CO(r^4)\ .
\end{equation}
we also compute that
\begin{equation}
\label{theA221} A_{2,21} = - \frac{1}{4} e^{2\lambda} h'\ .
\end{equation}
This is used in Section \ref{sec:expA}. One can solve for $f_2(z)$
and $g_1(z)$ from the EOMs \eqref{eqXA1}-\eqref{eqXA2}:
\begin{equation}
f_2(z) = - \frac{1}{4} e^\lambda \lambda'' \spa g_1(z) = -
\frac{1}{8} h'' - \frac{1}{4} h' \lambda'\ .
\end{equation}
By going order by order one can similar find $f_3(z)$ and
$g_2(z)$, and then $f_4(z)$ and $g_3(z)$, and so on. Therefore,
once $h(z)$ and $\lambda(z)$ are given, $X(r,z)$ and
$\tilde{A}(r,z)$ are uniquely determined.

\end{appendix}

\addcontentsline{toc}{section}{References}


\providecommand{\href}[2]{#2}\begingroup\raggedright\endgroup

\end{document}